\documentclass [prb,twocolumn,notitlepage,superscriptaddress,floatfix] {revtex4-1}
\usepackage{amsmath}
\usepackage{graphicx}
\usepackage{lmodern}
\usepackage{amsmath}

\usepackage{color}
\usepackage{amssymb}
\newcommand{\beq} {\begin{equation}}
\newcommand{\eeq} {\end{equation}}
\newcommand{\bea} {\begin{eqnarray}}
\newcommand{\eea} {\end{eqnarray}}
\newcommand{\be} {\begin{equation}}
\newcommand{\ee} {\end{equation}}
\renewcommand{\(}{\left(}
\renewcommand{\)}{\right)}
\renewcommand{\[}{\left[}
\renewcommand{\]}{\right]}

\DeclareMathOperator{\Tr}{Tr}

\newcommand{\K} {{\bf K}}

\begin{document}

\title {Fluctuating charge order in the cuprates: spatial anisotropy and feedback from superconductivity}
\author{Yuxuan Wang}
\affiliation{Department of Physics, University of Wisconsin, Madison, WI 53706, USA}
\affiliation{Department of Physics and Institute for Condensed Matter Theory,
University of Illinois at Urbana-Champaign, 1110 West Green Street, Urbana, Illinois, 61801, USA}
\author{Debanjan Chowdhury}
\affiliation{Department of Physics, Harvard University, Cambridge, MA 02138, USA}
\author{Andrey V. Chubukov}
\affiliation{William I. Fine Theoretical Physics Institute,
and School of Physics and Astronomy,
University of Minnesota, Minneapolis, MN 55455, USA}
\date{\today}

\begin{abstract}
We analyze the form of static charge susceptibility $\chi (q)$ in underdoped cuprates near axial momenta $(Q,0)$ and $(0,Q)$ at which short-range static charge order
has been observed. We show that the momentum dependence of $\chi (q)$ is anisotropic, and
the correlation length in the longitudinal direction is larger than in the transverse direction.  We show that
correlation lengths in both directions
decrease once the system evolves into a superconductor, as a result of the competition between superconductivity and charge order. These results are in agreement with resonant x-ray scattering data [R. Comin et al., Science {\bf 347}, 1335 (2015)].
We
 also argue that
 density and current components of the charge order parameter are affected differently by superconductivity --  the
  charge-density component is reduced
    less than the current component
     and hence  extends deeper into the superconducting state. This gives rise to two distinct charge order transitions at zero temperature.
 \end{abstract}
\maketitle

{\it Introduction.-}~
Understanding charge order (CO) in high-$T_c$  cuprates and its interplay with superconductivity is essential for the  understanding of the complex phase diagram of these materials.  An incommensurate charge order, accompanied by spin-order,
 was originally discovered in La-based cuprates~\cite{tranquada,tranquada1}, but recently was also found to occur in Y- Bi-, and Hg- based materials~\cite{ybco,ybco_2,ybco_1,x-ray, x-ray_1,davis_1,mark_last}, without
  an
  accompanying spin-order.
 A true long-range charge order has so far been observed only
 in a finite magnetic field \cite{mark_last}, but short-range static order (probably pinned by impurities) has been detected already in zero field.
The CO has an axial momentum ${\bf Q}=Q_y=(0,Q)$ or $Q_x=(Q,0)$ with $Q\sim (0.2-0.3)\times2\pi$,
  and can potentially be uni-axial (stripe), with only $Q_x$ or $Q_y$  within a  given domain, or bi-axial (checkerboard) with $Q_x$ and $Q_y$ present in every domain.  Recent STM and x-ray experiments~\cite{davis_15,comin_1} point towards the uni-axial order, at least at small values of the doping.
  The CO is often termed as  charge-density-wave (CDW) to emphasize that it develops with incommensurate momenta, although its on-site component is subleading to bond component because the measured form-factor for CO has predominantly a $d$-wave form~\cite{davis_1}.

The origin of the CO
 is still a subject of intense debates~\cite{ms,efetov, laplaca,charge,patrick,pepin,debanjan,flstar,kivelson,atkinson,agterberg,we_daniel,charge_new}.
Within one scenario, the charge order is induced by soft antiferromagnetic fluctuations \cite{ms,efetov, laplaca,charge}. This fits into the generic scenario that antiferromagnetism is the primary order parameter (i.e., the one whose fluctuations develop already at ``high" energies, comparable to the bandwidth), while
 CDW, its cousin pair-density-wave (PDW), and uniform $d$-wave superconductivity all develop as secondary orders induced by soft, low-energy magnetic fluctuations before the system becomes magnetically ordered. In another scenario, charge order is induced by lattice vibrations \cite{phonon}; in this case lattice and electronic degrees of freedom should be taken into account on equal footing.
  And in yet another scenario,
   CO
    emerges in the process of the system
     transformation from a conventional metal to a Mott insulator \cite{mott}. If charge order reflects the crossover towards Mott physics, then the tendency towards localization of electronic states cannot be neglected even above optimal doping.

 A
 way to
  distinguish between  these scenarios is to use the existing experimental data, particularly the ones for which the data analysis
   does not involve fitting parameters.  Recent x-ray scattering data in underdoped YBCO, reported in ~Ref.\ \onlinecite{comin_1}, can be used for this purpose.
   The data shows that the momentum structure of the charge
   susceptibility $\chi (q)$ near $Q_x$ and $Q_y$ is anisotropic, and
    the longitudinal correlation length is larger than the transverse one, i.e. if $\chi (q)$ at  ${\bf q} = {\bf Q} + \tilde{\bf q}$ is approximated by a Lorentzian $\chi^{-1} (q) \sim \xi^{-2} + A^2_{\parallel} {\tilde q}^2_{\parallel} + A^2_{\perp} {\tilde q}^2_{\perp}$, then
    $A_{\parallel} > A_\perp$.

  In this paper, we verify whether this condition is reproduced within the two itinerant scenarios -- the magnetic one and the phonon one. Exploring the Mott scenario is beyond the scope of this work.

    The charge susceptibility of itinerant electrons, $\chi (q)$, is generally related by a RPA-type formula to the static particle-hole polarization bubble between low-energy fermions separated by ${\bf q}$ (``hot" fermions).
     The difference between the two scenarios is that in the magnetic one the interaction is peaked at momentum $\K=(\pi,\pi)$ and connects fermions from two different hot regions. Specifically, magnetic interaction moves the center of mass momentum of a pair of hot fermions  from ${\bf k}_0$ to ${\bf k}_\pi = {\bf k}_0 + \K$. Then, one needs to apply spin-fluctuation mediated scattering twice to move fermions back to the same hot region (see Fig.\ \ref{fig:1}), and, as a consequence,  $\chi^{-1} (q) \propto 1 - U^2 \Pi_{k_0} ({\bf q}) \Pi_{k_\pi} ({\bf q})$, where $\Pi_{k_0} ({\bf q})$ and $ \Pi_{k_\pi} ({\bf q})$ are polarization operators made out of hot fermions with relative momentum ${\bf q}$ and center of mass momentum near $k_0$ and $k_\pi$, respectively (see Fig.\ \ref{fig:2}a).   In the phonon scenario, the interaction acts independently within each hot region, and $\chi^{-1} (q) \propto 1 - {\bar U} [\Pi_{k_0} ({\bf q}) + \Pi_{k_\pi} ({\bf q})]$.  In both cases, the momentum dependence of $\chi (q)$ is solely determined by  the polarization bubbles and does not depend on the strength of the interaction $U$  or ${\bar U}$.

       The polarization bubbles $\Pi_{k_0} ({\bf q})$ and
        $\Pi_{k_\pi} ({\bf q})$ depend on the Fermi surface geometry in the vicinity of the hot spots and also on the choice of the upper cutoff
 in the momentum deviations from the hot spots.
  The presence of the cutoff, $\Lambda$, reflects the fact that the  interaction in the charge channel can be approximated by a constant ($U$ or ${\bar U}$) only
  in  a finite range around a hot spot, outside of which it drops rapidly. In particular, within spin-fluctuation scenario, $\Lambda$ depends on the
    distance to the magnetic QCP -- it tends to a constant at the QCP and scales as inverse magnetic correlation length $\xi^{-1}_s$, when $\xi_s$ drops below a certain value~\cite{charge_new,yuxuan_last}.

 We
 report the results of analytic computations of $\Pi_{k_0} (q)$ and $\Pi_{k_\pi} (q)$
  using a hard cutoff (to be defined below).  In the magnetic scenario,
  the longitudinal charge correlation length turns out to be larger than the transverse one, in agreement with the data~\cite{comin_1}.  For the phonon scenario, the result is the opposite -- the transverse correlation length is larger.  Taken at face value, this observation selects magnetic mechanism of CO formation
   over the phonon one, although it indeed does not preclude phonon scattering as a subleading mechanism of CO.

We also consider how charge susceptibility $\chi (q)$ gets modified once the system becomes a $d$-wave superconductor.  We find that the key effect of superconductivity is the reduction of $\Pi_{k_0} ({\bf Q})$ and $\Pi_{k_\pi} ({\bf Q})$  due to competition between CO and superconducting order parameters. As a result, both longitudinal and transverse CO correlation lengths get smaller in the superconducting state.  A more subtle result is that the reduction of the bubble is different for density and current components of CO.  The two are symmetric and antisymmetric combinations of
the
 incommensurate charge order parameters
 $\Delta^Q_k \equiv c^{\dagger}_{k+Q/2, \alpha} \delta_{\alpha\beta}c_{k-Q/2,\beta}$
 with $k=\pm k_0$.
 The current component changes sign under time-reversal and once it develops along with the density component, the CO spontaneously breaks time-reversal symmetry~\cite{charge,rahul}.  In the normal state, density and current susceptibilities are equal [both are $\chi (q)$], as long as ${\bf k}_0$ and $-{\bf k}_0$ are well separated such that one can neglect bilinear coupling between $\Delta^Q_{k_0}$ and $\Delta^Q_{-k_0}$.
  In the superconducting state,
   $\chi^{-1} ({\bf Q})$ for the current component of CO gets shifted by  $\Delta_{sc}^2/(T \Lambda v_F)$ due to negative feedback from long-range superconducting order, $\Delta_{sc}$. For the density component of CO, such
   term cancels out and the shift is much smaller, of order $\Delta_{sc}^2/(\Lambda v_F)^2
   \times \log(\Lambda/T)$. As a result, the
density component of CO is
 much less
 affected by superconductivity  and should persist
 deeper into the superconducting state. This is in agreement with the x-ray data, which found that the measured charge density fluctuations persist down to lowest temperatures~\cite{comin_1}, deep into the superconducting state, where they also likely get pinned by quenched disorder \cite{nie}.
  Likewise, at
   zero temperature, the CO emerging from pre-existing superconducting state
   should initially have no current component.

\begin{figure}
\includegraphics[width=0.7\columnwidth]{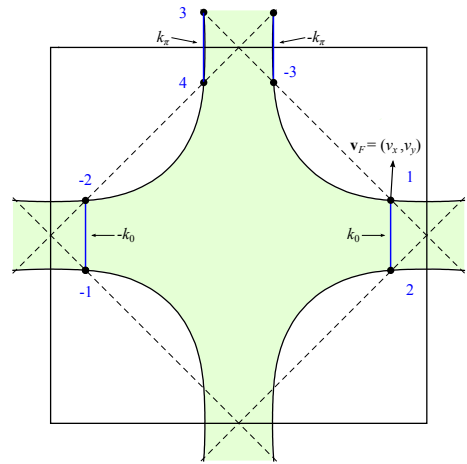}
\caption{The Fermi surface (black) with the occupied states shown in green. The location of the hot spots for CO (the points on the Fermi surface  separated by $Q_y=(0,Q)$ or $Q_x=(Q,0)$) are marked as $1, 2, -1, -2,...$ with the corresponding directions of Fermi velocities, $v_F$.  The notations $\pm k_0$ and $\pm k_\pi$ are for the center of mass momentum of charge order
  parameter  $\Delta^Q_k = c^{\dagger}_{k+Q/2, \alpha} \delta_{\alpha\beta}c_{k-Q/2,\beta}$.}
\label{fig:1}
\end{figure}

\begin{figure}
\includegraphics[width=0.7\columnwidth]{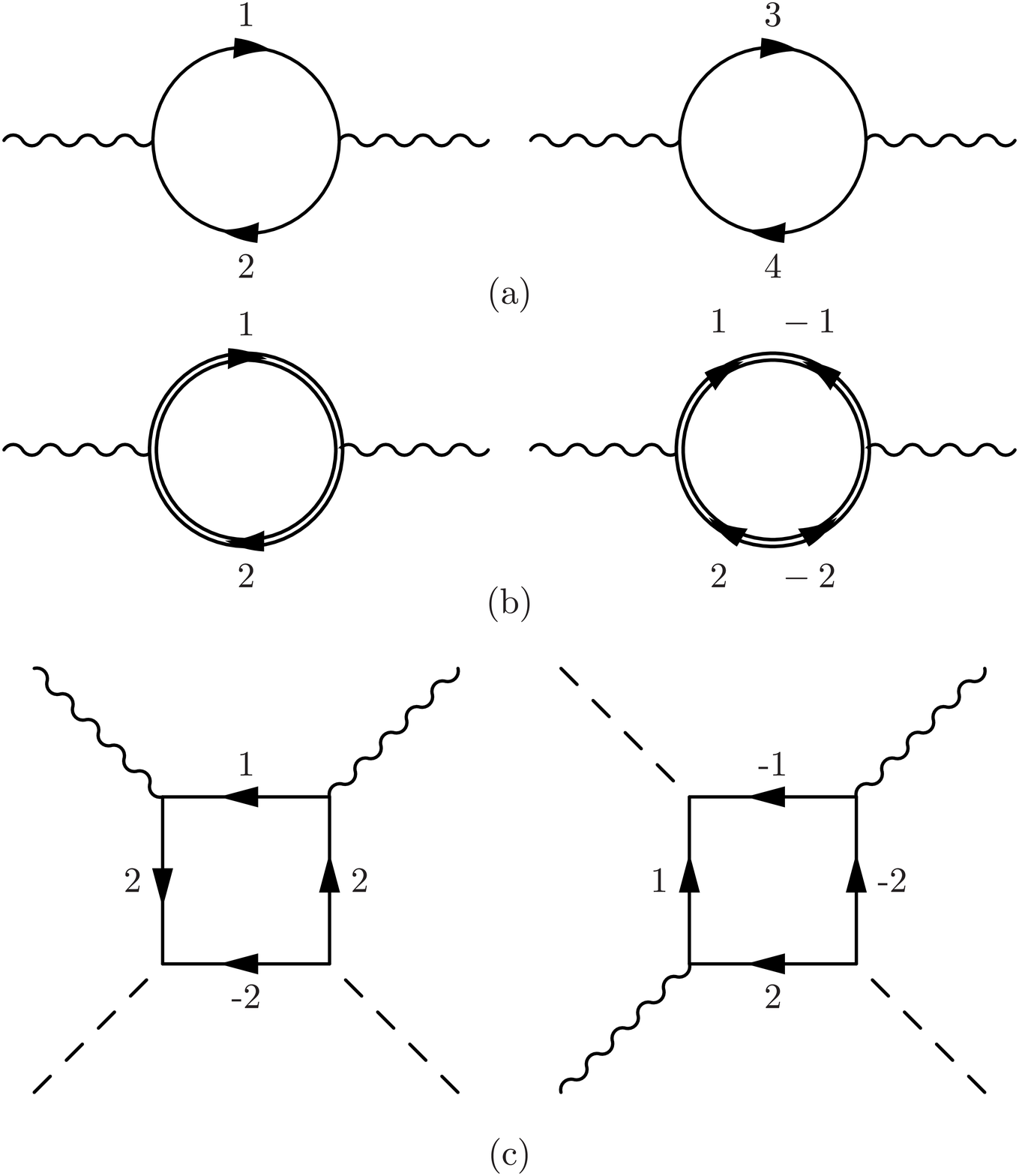}
\caption{The Feynman diagrams for (a) $\Pi_{k_0}(Q)$ and $\Pi_{k_\pi}(Q)$ in the normal state, and, (b) normal and anomalous contributions to $\Pi_{k_0}(Q)$ in the superconducting state.  (c) Four-point diagrams contributing to $\beta_{11}$ and $\beta_{12}$ in the free-energy (Eqn.\ref{FE}) in the normal state.}
\label{fig:2}
\end{figure}

{\it Polarization operators in the normal state.-}~ For the computation of polarization operators we used the Fermi surface shown in Fig.\ \ref{fig:1}. For definiteness we set ${\bf Q} = {\bf Q}_y = (0,Q)$. The results for ${\bf Q} = {\bf Q}_x$ are identical by a $\pi/2$ rotation. There are
   four ``hot" regions in the Brillouin zone. For the two regions with ${\bf k}_0 = (\pi-Q/2,0)$ and $-{\bf k}_0$, the Fermi velocity of the two fermions separated by ${\bf Q}$ are almost anti-parallel, while for the other two, with ${\bf k}_\pi = (-Q/2, \pi)$ and $-{\bf k}_\pi$, the velocities are nearly parallel. In the notations in Fig. \ref{fig:1}, this implies $v_y \gg v_x$ ($v_y/v_x \approx 13.6$ in BSCCO, see Ref.~\onlinecite{mike}).

 We impose hard cutoff by requiring that momentum of each fermion in the particle-hole bubble $\Pi_{k_0} (q)$ and $\Pi_{k_\pi} (q)$
       differs from the corresponding  hot spot by no more than $\Lambda$ by amplitude.  In the hot region with center of mass momentum of a pair near $k_0$ we define ${\bf k}_{1,2} = {\bf k}_0 \pm ({\bf Q}_y +  \tilde{\bf q})/2+ \tilde {\bf k}$, and the condition reads $|\tilde {\bf k} \pm {\tilde {\bf q}}/2| < \Lambda$.
        The analogous condition holds in the hot region with center of mass momentum of a pair near $k_\pi$.
    We assume that $\Lambda$ is small compared with inverse lattice spacing, in which case we can expand the dispersion of a hot fermion to linear order in deviation from a hot spot.   Under these condition, we obtained analytical expressions
     for the  polarization bubbles  $\Pi_{k_0} (q)$ and $\Pi_{k_\pi} (q)$ to leading order in $\tilde {\bf q} = {\bf q} - {\bf Q}_y$.
      The calculation is lengthy but straightforward \cite{SI}, and the result is
     \bea
      \Pi_{k_0}(\tilde {\bf q}) &=A v_x\left(2\Lambda  \log{\frac{v+ v_y}{v_x}} - |{\tilde q}_y| \sin^{-1}{\frac{v_y}{v}}- |{\tilde q}_x|  \log{\frac{v}{v_x}} \right),\nonumber \\
      \Pi_{k_\pi}(\tilde {\bf q}) &= A v_y\left(2\Lambda  \log{\frac{v+ v_x}{v_y}} - |{\tilde q}_y| \sin^{-1}{\frac{v_x}{v}}- |{\tilde q}_x| \log{\frac{v}{v_y}} \right) \nonumber\\
      \eea
       where $v= \sqrt{v^2_x + v^2_y}$ and  $A = 1/(\pi^2 v_x v_y)$. It is interesting to note that the singular terms proportional to $|{\tilde q}_{x,y}|$ are independent of the cutoff, $\Lambda$.

     The case of hard cutoff is
       somewhat
       special because the expansion of polarization bubbles in $\tilde {\bf q}$ is non-analytic and holds in powers of $|{\tilde q}_x|$ and $|{\tilde q}_y|$ rather than in ${\tilde q}^2_x$ and  ${\tilde q}^2_y$.  {\it We verified that the quadratic dependence emerges immediately once we soften the cutoff.}  Still, even for strictly hard cutoff one can analyze the anisotropy of the inverse charge susceptibility $\chi^{-1} (q)$  by comparing the prefactors for longitudinal ($|{\tilde q}_y|$) and transverse ($|{\tilde q}_x|$) momentum dependencies.  By continuity, the anisotropy should survive upon softening of the cutoff.

     To analyze the anisotropy, we use $v_y \gg v_x$ and expand $ \Pi_{k_0}$ and  $\Pi_{k_\pi}$ in small $v_x/v_y$ limit \cite{SI}.
     In the magnetic scenario
      we find $\chi^{-1} (q) \propto C_0 + C_y |{\tilde q}_y| + C_x |{\tilde q}_x| + ...$, where
       $C_0 = 1/U^2 -4 A^2 \Lambda^2 v^2_x \log{2 v_y/v_x}$,  $C_y = 2 A^2 \Lambda v^2_x (\log{2v_y/v_x} + \pi/2), C_x = 2 A^2 \Lambda v^2_x \log{v_y/v_x}$ and the ellipses denote higher order terms in ${\tilde q}_{x,y}$.  We remind that ${\bf q} = {\bf Q}_y + \tilde {\bf q}$, hence ${\tilde q}_y$ is longitudinal component.
 Taking the ratio $C_y/C_x$ we immediately see that $C_y/C_x = 1 + [(\pi/2 + \log{2})/\log{(v_y/v_x)}] >1$,  i.e., the effective correlation length $\xi_{\parallel} = C_y/C_0$ is larger than  $\xi_{\perp} = C_x/C_0$.  This is consistent with the data~\cite{comin_1}.
For $v_y/v_x \approx 13.6$, we obtained, without expanding, $\xi_{\parallel}/\xi_{\perp} = 1.87$, which is reasonably close to the experimental ratio of around 1.5.

  In the phonon scenario
   we obtain $\chi^{-1} (q) \propto {\bar C}_0 + {\bar C}_y |{\tilde q}_y| + {\bar C}_x |{\tilde q}_x| + ...$, where  now, to logarithmic accuracy,  ${\bar C}_y/{\bar C}_x = (\pi/2+1)/\log{(v_y/v_x)}$.  Then ${\bar C}_y/{\bar C}_x$ is smaller than one, at least when $v_y/v_x$ is large enough. For $v_y/v_x \approx 13.6$, we obtained, without expanding, $\xi_{\parallel}/\xi_{\perp} =0.98$.

We also computed $\Pi_{k_0} (\tilde {\bf q})$ and $\Pi_{k_\pi}(\tilde {\bf q})$ numerically for a
specific Lorentzian cutoff,
which we imposed by
inserting into the integrands for the bubbles an additional factor, $\Lambda^2/(\Lambda^2 + (\tilde {\bf k} + \tilde {\bf q}/2)^2) \times \Lambda^2/(\Lambda^2 + (\tilde {\bf k} - \tilde {\bf q}/2)^2)$, but not restricting integration over momentum.  One can immediately make sure that in this case the expansion in ${\tilde q}$  holds in powers of ${\tilde q}^2$.
 We again find that in a magnetic scenario $\xi_{\parallel}/\xi_{\perp} >1$. However this ratio is much larger. From this perspective, the hard cutoff gives better agreement with the data.

{\it Superconducting state.-}~  The polarization operators $\Pi_{k_0} (q)$ and $\Pi_{k_\pi} (q)$ in the superconducting state are obtained in
 a conventional way, by combining bubbles made out of normal and anomalous fermionic Green's functions
 (Fig. \ref{fig:2}b). The full expressions are more involved and we didn't obtain analytical formulas even for hard cutoff.  In general, both $\Pi_{k_0} ({\bf Q}_y)/\Pi_{k_\pi} ({\bf Q}_y)$ and the momentum-dependent terms  in the polarization operators evolve with the superconducting gap $\Delta_{sc}$.  The effect, however, is stronger for $\chi ({\bf Q}_y)$ rather than for ${\tilde q}$-dependent terms because for $\chi ({\bf Q}_y)$ superconductivity-induced shift has to be compared with the initially small value of the mass of the charge susceptibility at ${\bf Q}_y$.  We therefore focus on the renormalization of the polarization operators right at ${\bf q} = {\bf Q}_y$.  The calculations, which we describe in more detail below, show expected trends -- superconductivity competes with CO, and once long-range superconducting order develops, it tends to delay the appearance of CO.  This effect is very typical for competing orders and has been  recently discussed in detail for Fe-pnictides~\cite{Fe-competition}.
  Because $\chi^{-1} ({\bf Q}_y)$
  increases, both longitudinal and transverse charge correlation lengths go down.
     The data~\cite{comin_1}  show the same trend.

 A more subtle issue is the magnitude of superconductivity-induced shift. Near $T_c$ (i.e., for relatively small $\Delta_{sc}$), the shift originates from $\beta_{ij} |\Delta_{sc}|^2 \Delta^Q_{k_i} (\Delta^Q_{k_j})^*$ terms in the Free energy, where, we remind,  $\Delta^Q_k$ is fluctuating CO field (not the condensate), and $k_i$ is either $\pm k_0$ or $\pm k_\pi$. Since the superconducting pair has zero total momentum, it couples to fermions located only  within one corner of the Brillouin zone; hence coupling terms for $\pm k_0$ and $\pm k_\pi$ can be considered separately. At the same time, superconductivity pairs fermions with opposite momenta, hence both $|\Delta^Q_{k}|^2$ and $\Delta^Q_k (\Delta^Q_{-k})^*$ couple to $|\Delta_{sc}|^2$.

 For definiteness, let's set, as before,
  ${\bf Q} = {\bf Q}_y$ and focus on the region where center of mass momentum of charge order parameter is $\pm k_0$.
 The Free energy is
 \begin{widetext}
  \beq
 F= \chi^{-1} (Q_y) \left( |\Delta^{Q_y}_{k_0}|^2 + |\Delta^{Q_y}_{-k_0}|^2 \right)  +  |\Delta_{sc}|^2 \left[\beta_{11} \left(|\Delta^{Q_y}_{k_0}|^2 + |\Delta^{Q_y}_{-k_0}|^2\right) + \beta_{12}
  \left(\Delta^{Q_y}_{k_0} (\Delta^{Q_y}_{-k_0})^* + (\Delta^{Q_y}_{k_0})^* \Delta^{Q_y}_{-k_0}\right)\right]
  \label{FE}
\eeq
\end{widetext}
 (We neglected spatial fluctuations of CO and terms unrelated to our purposes.)
 This Free energy is easily diagonalized by introducing $\Delta_d = (\Delta^{Q_y}_{k_0} + \Delta^{Q_y}_{-k_0})/\sqrt{2}$ and $\Delta_c =  (\Delta^{Q_y}_{k_0} - \Delta^{Q_y}_{-k_0})/\sqrt{2}$. In terms of these variables
 \bea
 F &&= |\Delta_d|^2 \left(\chi^{-1} (Q_y) + 2 |\Delta_{sc}|^2 (\beta_{11} + \beta_{12})\right) \nonumber \\
 && + ~|\Delta_c|^2 \left(\chi^{-1} (Q_y) + 2  |\Delta_{sc}|^2(\beta_{11} - \beta_{12})\right).
 \eea
We see that the shift of $\chi^{-1} (Q_y)$ due to superconductivity  is generally different for
symmetric density  and antisymmetric current
 components of CO ($\Delta_d$ and $\Delta_c$, respectively).

 The couplings $\beta_{11}$ and $\beta_{12}$ can be evaluated either by using Hubbard-Stratonovich (HS) formalism, or by expanding particle-hole bubbles to order $\Delta_{sc}^2$.  In the HS formalism, these two terms are given by square diagrams made out of four fermionic Green's functions in the normal state (shown in Fig. \ref{fig:2}c).

   We computed $\beta_{11}$ and $\beta_{12}$ for a model with hard cutoff and found that the dominant piece in each is a cutoff-independent
   term, which scales as $1/T$.  At the lowest $T$, the expansion in powers of $\Delta_{sc}$ does not hold, and $1/T$ divergence is cut by $1/|\Delta_{sc}|$.
    Upon a more careful look, we found that universal (i.e., cutoff-independent)  $1/T$ terms in $\beta_{11}$ and $\beta_{12}$ come with exactly
     opposite coefficients, i.e., $1/T$ terms cancel out in $\beta_{11} + \beta_{22}$.  This cancellation has not been noticed before.
     From mathematical perspective, this cancellation is similar to the cancellation between vertex and self-energy corrections to uniform density-density correlator. In the latter case, however, the cancellation is exact and it enforces Ward identity associated with particle number conservation.  In our case, the cancellation is not required
 by symmetry and is not exact -- only the leading $1/T$ terms cancel out in $\beta_{11} + \beta_{12}$. The subleading terms  do not cancel.
 These subleading terms
 are, however, much smaller,  and the correction to $\chi^{-1} (Q_y)$ from the superconducting order  is given by
  $\alpha |\Delta_{sc}|^2/\Lambda\times \log(\Lambda/T)$
  ($\alpha >0$).
The outcome is that the mass of the density component of charge order parameter (the one measured by x-ray) goes up in the presence of superconductivity (and the correlation length, which scales as inverse mass, goes down); however this effect is
  small. A small coupling between competing charge density  and superconducting orders also implies that the order which appears first does not rapidly destroy the other one, hence CDW and superconductivity co-exist over a sizable range of dopings.

\begin{figure}
\includegraphics[width=0.8\columnwidth]{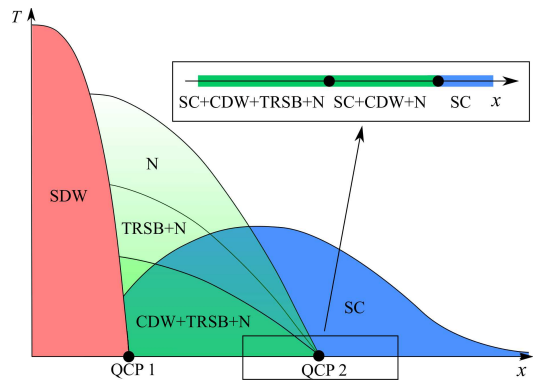}
\caption{The phase diagram in variables $T$ and $x$ (adapted from Ref. \onlinecite{chargeYW}), with more details
near $T=0$. 
   N and TRSB stand for nematic and time-reversal symmetry breaking, respectively,
  and CDW is the phase with broken translational symmetry (in a clean system). }
\label{fig:3}
\end{figure}

  At the same time, for the current CO component $\Delta^Q_c$,  $\Delta_{sc}^2/T$ contributions in $\beta_{11}$ and $\beta_{12}$ add up, i.e., for this component superconductivity has  stronger negative impact.    In particular,  when  CO emerges inside the superconducting dome, only its density component becomes non-zero when $-\chi^{-1} (Q) =  |\Delta_{sc}|^2 (\beta_{11} + \beta_{22}) \sim  |\Delta_{sc}|^2/\Lambda$.  Current component emerges at smaller dopings, when (and if) the condition
 $-\chi^{-1} (Q) =  |\Delta_{sc}|^2 (\beta_{11} - \beta_{22}) \sim  |\Delta_{sc}|^2/(\text{max} (T,|\Delta_{sc}|))$ is satisfied, and
  co-exists  with superconductivity in a narrower range.
 We show this in Fig.\ \ref{fig:3}.  Since the current component of CO is responsible for the breaking of time-reversal symmetry (TRS), its absence over some doping range where the density component of CO is present implies that in this range CO cannot be the source of TRS breaking. Alternatively speaking, if TRS breaking is caused by charge order, the end point of TRS breaking transition should end up at $T=0$ at a smaller doping than the onset of CO.  It would be interesting to test this in
  Kerr and
elastic neutron scattering measurements in the superconducting state~\cite{kerr,greven,sidis}.

 {\it Summary.-}~In this paper we considered three aspects associated with uni-axial charge order in the underdoped cuprates.  First, we analyzed the anisotropy of the charge order correlation length in the normal state, detected in recent x-ray measurements. Our goal was to investigate whether
   these data allow one to distinguish between magnetic and phonon-based mechanisms of CO formation.  We argued that the magnetic scenario yields results consistent with the data in Ref.\ \onlinecite{comin_1}. Second, we argued that both longitudinal and transverse charge correlation lengths decrease in the presence of a true superconducting order, primarily because this order increases the mass of the charge order propagator. Finally, we found that the mass increase is different for the density and the current components of CO (symmetric and antisymmetric components with respect to the flip of a center of mass momenta of fermions which form CO).  The mass increase is strong for the current component and is parametrically weaker for the density component. As a result, under the umbrella of superconductivity, the density component of CO exists in a wider range of doping compared to the current component. Since the current component of CO is responsible for TRS breaking, we propose to use this fact to test whether TRS breaking is associated with incommensurate charge order.

We thank E. Berg, R. Fernandes and S. Sachdev for fruitful discussions. The work was supported by the NSF-
DMR-1523036 (YW and AC) and
DMR-1360789 (DC).
 DC acknowledges the
 hospitality of FTPI, Minneapolis and
  PITP, Waterloo, where a part of this work was completed. AC and DC acknowledge hospitality of MPI-PKS, Dresden. \emph{}

\begin{widetext}
\maketitle
\begin{center} \textbf{Supplementary Material} \end{center}
\section{Evaluation of the particle-hole bubbles in the normal state}

In this section we evaluate the particle-hole bubble $\Pi_{k_0}$ and $\Pi_{k_\pi}$ associated with the incommensurate charge order with momentum around, say, ${\bf Q}=Q_y$. The results obtained below can be trivially extended to the case for ${\bf Q}=Q_x$ by rotational symmetry.

By simple dimensional analysis the particle-hole bubble $\sim \int d\omega d^2 k~ GG$, where $G$ is the Fermionic Green's function in the normal state, scales as the upper momentum cutoff $\Lambda$. For the spin fluctuation scenario, the cutoff scheme in momentum is natural since the structure of the interaction function dictates that only fermions in the hot region are relevant. For a small antiferromagnetic correlation length $\xi_s$, this cutoff around corresponding hot spots is given by roughly $\Lambda\sim \xi_{s}^{-1}$. On the other hand, for the phonon scenario, it is not clear how to impose the momentum cutoff through microscopic mechanism. However, such a cutoff near hot regions is still phenomenologically required -- otherwise the local minima of $\Pi({\bf Q})$ will be located at $\bf Q$'s connecting the nested portions of the Fermi surface near the antinodes, rather than the experimentally-detected~\cite{comin} ${\bf Q}=Q_y$. To ensure an unbiased comparison between the two scenarios, we confine the typical momentum deviation from hot spots to be of order $\Lambda$, without specifying the microscopic origin of $\Lambda$. We adopted both ``hard" and ``soft" cutoff schemes, and found that a hard cutoff yields a better match with the experimental data.

\subsection{Particle-hole bubbles with a hard cutoff}
We now evaluate the bubble near hot spots 1 and 2 with a hard momentum cutoff.
\begin{align}
\Pi_{k_0}({\bf Q}+{\bf q})=-T\sum_m\int_{|{\bf k}\pm {\bf q}/2|<\Lambda} \frac{d^2k}{(2\pi)^2}\frac{1}{i\omega_m-\epsilon_1({\bf k}+{\bf q}/2)}\frac{1}{i\omega_m-\epsilon_2({\bf k}-{\bf q}/2)}.
\label{eq:3}
\end{align}
We first evaluate this integral for ${\bf q}=q_x$.   Performing the summation in frequency we have,
\begin{align}
\Pi_{k_0}(q_x)=&-\int_{|{\bf k}\pm {\bf q}/2|<\Lambda}\frac{d^2k}{4\pi^2}\frac{f(\epsilon_1(k+q_x/2))-f(\epsilon_2(k-q_x/2))}{\epsilon_1(k+q_x/2)-\epsilon_2(k-q_x/2)},\nonumber\\
=&\frac{1}{4\pi^2}\int_C\frac{d^2k}{|2v_yk_y+v_xq_x|}.
\label{eq:2}
\end{align}
where $f(\epsilon)$ is the Fermi function. In the last step we have approximated $f(\epsilon)$ by the step function, and we have used the linear dispersion $\epsilon_{1,2}(k)=v_xk_x\pm v_yk_y$. The integration region $C$ is given by $\{\epsilon_1(k+q_x/2)\epsilon_2(k-q_x/2)<0\} \cap \{|{\bf k}\pm {\bf q}/2|<\Lambda\}$, which we depict in Fig.\ \ref{rega}(b) for the case $v_x|q_x|\ll \Lambda$. For comparison we also show the integration region for ${\bf q}=0$ case in Fig.\ \ref{rega}(a).

We can then divide the integration region into more regular shapes and evaluate them separately. Before we do so, it is convenient to introduce rescaled variables $y=2v_y k_y/(v_xq_x)$, $x=2k_x/q_x$, and $Q=2v_y\Lambda/(v_x|q_x|)$:
\begin{align}
\Pi_{k_0}(q_x)=\frac{|q_x|}{16\pi^2v_y}\int_{C'}\frac{dx~dy}{|y+1|},
\label{eq:1}
\end{align}
where $C'$ is now given by $\{|x|<|y+1|\}\cap \{|x|<\frac{v_x}{v_y}\sqrt{Q^2-y^2}-1\}$. We take the limit $|q_x|\ll \Lambda$, and hence $Q\gg 1$. By dimensional analysis, the integral from Eq.\ (\ref{eq:1}) in leading order scales as $Q$, which, when combined with the prefactor $|q_x|/(16\pi^2v_y)$ gives a contribution $\sim\Lambda$. This is nothing but $\Pi_{k_0}(\bf Q)$. To obtain the leading order correction for $\bf q=q_x$, one needs to keep contributions up to next leading order in $Q$.

Using Fig.\ \ref{rega}(b), we can divide the integration region $C'$ as $\Pi_{k_0}(q_x)={|q_x|}/({16\pi^2v_y})\(\int_{C_1'}+\int_{C_2'}+\int_{C_3'}\)\equiv{|q_x|}/({16\pi^2v_y})( I_1+I_2+I_3)$, where
\begin{align}
I_1=&\int_{C_1'}\frac{dx~dy}{|y+1|}=\int^{Qv_x/v-2v_y^2/v^2}_{-Qv_x/v}dy\int_{-|y+1|}^{|y+1|}dx \frac{1}{|y+1|} \nonumber\\
I_2=&\int_{C_2'}\frac{dx~dy}{|y+1|}=\int_{Qv_x/v-2v_y^2/v^2}^{\sqrt{Q^2-v_y^2/v_x^2}}dy\int_{-v_x\sqrt{Q^2-y^2}/v_y+1}^{{v_x\sqrt{Q^2-y^2}/v_y-1}} dx  \frac{1}{|y+1|} \nonumber\\
I_3=&\int_{C_3'}\frac{dx~dy}{|y+1|}=\int_{-\sqrt{Q^2-v_y^2/v_x^2}}^{-Qv_x/v} dy \int_{-v_x\sqrt{Q^2-y^2}/v_y+1}^{{v_x\sqrt{Q^2-y^2}/v_y-1}} dx  \frac{1}{|y+1|}.
\end{align}
The integration limits for $I_{1}$ comes from the condition $|y+1|<({v_x}/{v_y})\sqrt{Q^2-y^2}-1$ and has been expanded up to next leading order in $Q$.

\begin{figure}
\includegraphics[width=0.8\columnwidth]{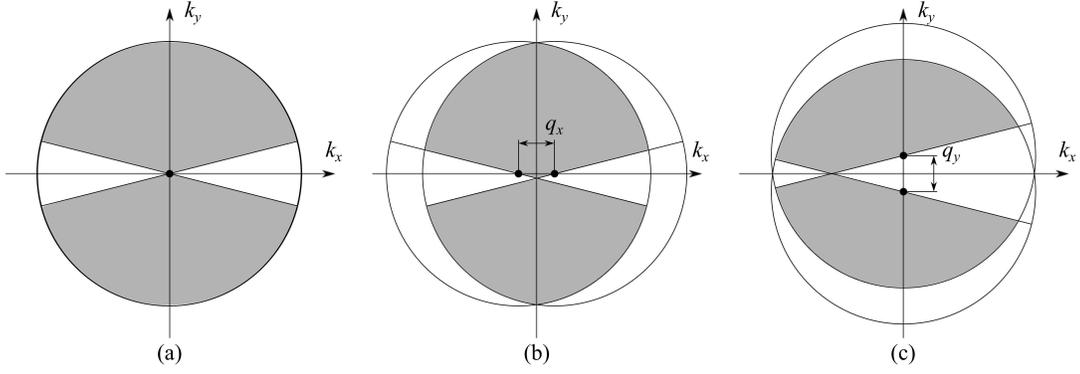}
\caption{The integration regions for particle-hole bubble near hot spots 1 and 2 with (a) total momentum right at ${\bf Q}$, (b) momentum deviation ${\bf q}=q_x$, and (c) momentum deviation ${\bf q}=q_y$. The circles centered around $\pm {\bf q}/2$ with radius $\Lambda$ are imposed by the hard upper momentum cutoff $|{\bf k}\pm {\bf q}/2|<\Lambda$, and the straight lines with slope $\pm v_x/v_y$ are FS's in the vicinity of hot spots 1 and 2.}
\label{rega}
\end{figure}

The evaluation of $I_1$ is trivial, which, to next leading order in large $Q$, gives 
\begin{align}
I_1=4Qv_x/v-4v_y^2/v^2.
\end{align} For $I_2+I_3$ we have
\begin{align}
I_2+I_3=&4\int_{Qv_x/v}^{\sqrt{Q^2-v_y^2/v_x^2}}\frac{y dy}{y^2-1}\(\frac{v_x}{v_y}\sqrt{Q^2-y^2}-1\)+2\int_{Qv_x/v-2v_y^2/v^2}^{Qv_x/v}\frac{dy}{y+1}\(\frac{v_x}{v_y}\sqrt{Q^2-y^2}-1\)\nonumber\\
\approx&4\int_{Qv_x/v}^{Q}\frac{dy}{y}\(\frac{v_x}{v_y}\sqrt{Q^2-y^2}-1\)+\(\frac{4v_y^2}{v^2}\)\frac{v}{Qv_x}\frac{v_x}{v_y}\sqrt{Q^2-\(\frac{v_xQ}{v}\)^2}\nonumber\\
=&\[\frac{4Qv_x}{v_y}\(-\frac{v_y}{v}+\log\frac{v+v_y}{v_x}\)-4\log\frac{v}{v_x}\]+\frac{4v_y^2}{v^2}.
\end{align}
In the second line above we have kept terms only up to next leading order in $Q$. Note, however, that the above result is {\it not} an approximation in the small $v_x/v_y$ limit but rather for an arbitrary $v_x/v_y$.

Combining the results for $I_{1,2,3}$, and using $\Pi_{k_0}(q_x)={|q_x|}/({16\pi^2v_y})( I_1+I_2+I_3)$, we find in original variables
\begin{align}
\Pi_{k_0}(q_x)=\frac{\Lambda}{2\pi^2 v_y}\log\frac{v+v_y}{v_x}-\frac{|q_x|}{4\pi^2v_y}\log\frac{v}{v_x}.
\label{qx}
\end{align}
The first term of Eq.\ (\ref{qx}) corresponds to $\Pi_{k_0}(q_x=0)\equiv \Pi_{k_0}({\bf Q})$. This term has been evaluated~\cite{debanjan} before, with a slightly different cutoff scheme. In the limit $v_y\gg v_x$, we have $\Pi_{k_0}(0)={\Lambda}/({2\pi^2v_y})\log (v_y/v_x)$, and this is fully consistent with previous result.

Next we evaluate the same bubble with momentum deviation ${\bf q}=q_y$ from the CDW momentum $\bf Q$. Similar to  Eq.\ (\ref{eq:2}) we start from
\begin{align}
\Pi_{k_0}(q_y)=&-\int_{|{\bf k}\pm {\bf q}/2|<\Lambda}\frac{d^2k}{4\pi^2}\frac{f(\epsilon_1(k+q_y/2))-f(\epsilon_2(k-q_y/2))}{\epsilon_1(k+q_y/2)-\epsilon_2(k-q_y/2)},\nonumber\\
=&\frac{1}{4\pi^2}\int_D\frac{d^2k}{|2v_yk_y|}.
\label{eq:4}
\end{align}
The integration region $D$ is given by $\{\epsilon_1(k+q_y/2)\epsilon_2(k-q_y/2)<0\} \cap \{|{\bf k}\pm {\bf q}/2|<\Lambda\}$, which we depict in Fig.\ \ref{rega}(c) for the case $v_y|q_y|\ll \Lambda$. Rescaling $\bar y=2k_y/(q_y)$, $\bar x=2v_xk_x/(v_yq_y)$, and $\bar Q=2v_x\Lambda/(v_y|q_y|)\gg1$, we rewrite Eq.\ (\ref{eq:4}) as
\begin{align}
\Pi_{k_0}(q_y)=\frac{|q_y|}{16\pi^2v_x}\int_{D'}\frac{d\bar x~d\bar y}{|\bar y|}.
\label{eq:5}
\end{align}
For simplicity of presentation, we drop the bars in the rescaled variables below.
From Fig.\ \ref{rega}(c) we see that the integration region $D'$ can be simply expressed as $|x+1|<|y|<(v_y/v_x)\sqrt{Q^2-x^2}-1$, where $Q\gg1$. We then integrate over $y$ first and obtain
\begin{align}
\Pi_{k_0}(q_y)=\frac{|q_y|}{8\pi^2v_x}\int_{-(v_y/v)Q
}^{(v_y/v)Q-2v_x^2/v^2}dx\log\frac{(v_y/v_x)\sqrt{Q^2-x^2}-1}{|x+1|},
\label{eq:10}
\end{align}
where the integration limits on $x$ is obtained by expanding the solution of $|x+1|=(v_y/v_x)\sqrt{Q^2-x^2}-1$ up to next leading order in $Q$. Evaluating Eq.\ (\ref{eq:10}) in the large $Q$ limit, we obtain
\begin{align}
\Pi_{k_0}(q_y)=&\frac{|q_y|}{8\pi^2v_x}\[\int_{-(v_y/v)Q}^{(v_y/v)Q}dx\log\frac{(v_y/v_x)\sqrt{Q^2-x^2}}{|x|}-
\int_{(v_y/v)Q-2v_x^2/v^2}^{(v_y/v)Q}dx\log\frac{(v_y/v_x)\sqrt{Q^2-x^2}}{|x|}\right.\nonumber\\
&\left.~~~~~~~~
-\int_0^{(v_y/v)Q}\frac{dx}{|x|}+\int_{-(v_y/v)Q}^{0}\frac{dx}{|x|}-\frac{v_x}{v_y}\int_{-(v_y/v)Q}^{(v_y/v)Q}\frac{dx}{\sqrt{Q^2-x^2}}\],
\end{align}
where the first term in the bracket is of order $Q$, which corresponds to $\Pi_{k_0}({\bf Q})$, and the last four terms are of order one. The second Evaluating the integrals, we obtain
\begin{align}
\Pi_{k_0}(q_y)=\frac{\Lambda}{2\pi^2 v_y}\log\frac{v+v_y}{v_x}
-\frac{|q_y|}{4\pi^2v_y}\sin^{-1}\frac{v_y}{v}.
\label{qy}
\end{align}
Combining Eqs.\ (\ref{qx},\ref{qy}), we get the full expression 
\begin{align}
\Pi_{k_0}({\bf Q}+{\bf q})=\frac{\Lambda}{2\pi^2 v_y}\log\frac{v+v_y}{v_x}-\frac{|q_x|}{4\pi^2v_y}\log\frac{v}{v_x}
-\frac{|q_y|}{4\pi^2v_y}\sin^{-1}\frac{v_y}{v}.
\label{qxy}
\end{align}

Finally, we move to the evaluation of $\Pi_{k_\pi}$, defined as the particle-hole bubble for fermions near hot spots 3 and 4. Similar to Eq.\ (\ref{eq:3}), we have
\begin{align}
\Pi_{k_\pi}({\bf Q}+{\bf q})=-T\sum_m\int_{|{\bf k}\pm {\bf q}/2|<\Lambda} \frac{d^2k}{(2\pi)^2}\frac{1}{i\omega_m-\epsilon_3({\bf k}+{\bf q}/2)}\frac{1}{i\omega_m-\epsilon_4({\bf k}-{\bf q}/2)}.
\label{eq:16}
\end{align}
We remind that $\epsilon_{3,4}=-v_yk_x\mp v_xk_y$, while $\epsilon_{1,2}=v_xk_x\pm v_yk_y$. Using the fact that the integrations over $k_x$ and $k_y$ are symmetric, we then see that $\Pi_{k_\pi}({\bf Q}+{\bf q})$ and $\Pi_{k_0}({\bf Q}+{\bf q})$ are simply related by $v_x\leftrightarrow v_y$, therefore
\begin{align}
\Pi_{k_\pi}({\bf Q}+{\bf q})=\frac{\Lambda}{2\pi^2 v_x}\log\frac{v+v_x}{v_y}-\frac{|q_x|}{4\pi^2v_x}\log\frac{v}{v_y}
-\frac{|q_y|}{4\pi^2v_x}\sin^{-1}\frac{v_x}{v}.
\label{qyx}
\end{align}
As a quick check, in the limit $v_x\ll v_y$, the $\bf q$ independent term becomes $\Pi_{k_\pi}({\bf Q})={\Lambda}/({2\pi^2 v_y})$. This is consistent with results previously obtained~\cite{debanjan}. Eqs.\ (\ref{qxy},\ref{qyx}) are the two main results of this Subsection.

When $v_y \gg v_x$ we can expand $ \Pi_{k_0}$ and  $\Pi_{k_\pi}$ in small $v_x/v_y$,
\bea
    \Pi_{k_0} (\tilde {\bf q})&\approx& A \left[2\Lambda v_x \log{\frac{2 v_y}{v_x}}  - |{\tilde q}_y| \frac{\pi}{2} v_x - |{\tilde q}_x| v_x \log{\frac{v_y}{v_x}} \right],  \nonumber \\
     \Pi_{k_\pi} (\tilde {\bf q}) &\approx& A \left[2\Lambda v_x - |{\tilde q}_y| v_x - |{\tilde q}_x| \frac{v^2_x}{2v_y} \right].
     \label{2}
 \eea

\subsection{Particle-hole bubbles with a soft cutoff}
Using a hard cutoff, we found that the leading order $\bf q$ dependence of the polarization bubbles has a non-analytical form $\sim|q|$. This $|q|$ dependence gets ``softened" to a $q^2$ one, if we use a soft upper cutoff instead of a hard one. For example, one can replace the integration limit $ ({\bf k} \pm {\bf q}/2)^2<\Lambda^2$ with a multiplier $\Lambda^{2N}/({\bf k} + {\bf q}/2)^N+\Lambda^N)/({\bf k} - {\bf q}/2)^N+\Lambda^N)$ and integrate over infinite range. This multiplier is fully analytical, hence the leading order $q$ dependence should be $\sim q^2$. For large $N$, this multiplier almost looks like a step function, which restores the hard cutoff. As a result, the coefficient of $q^2$ term increases, which diverges at $N\to \infty$ and gives a $\sim |q|$ behavior in the hard cutoff case. We have verified that the ratio between coefficients for $q_x^2$ and $q_y^2$ for a large but finite $N$ connects smoothly to the ratio between $|q_x|$ and $|q_y|$ coefficients obtained previously using a hard cutoff.

For completeness, in this subsection we perform a calculation of the bubbles for $N=2$, which is a Lorentzian cutoff. Namely, we define
\begin{align}
\Pi_{k_0}({\bf Q}+{\bf q})&=-T\sum_m\int \frac{d^2k}{(2\pi)^2}\frac{1}{i\omega_m-\epsilon_1({\bf k}+{\bf q}/2)}\frac{1}{i\omega_m-\epsilon_2({\bf k}-{\bf q}/2)}\frac{\Lambda^{4}}{[({\bf k} + {\bf q}/2)^2+\Lambda^2][({\bf k} - {\bf q}/2)^2+\Lambda^2]}\label{eq:6}\\
\Pi_{k_\pi}({\bf Q}+{\bf q})&=-T\sum_m\int \frac{d^2k}{(2\pi)^2}\frac{1}{i\omega_m-\epsilon_3({\bf k}+{\bf q}/2)}\frac{1}{i\omega_m-\epsilon_4({\bf k}-{\bf q}/2)}\frac{\Lambda^{4}}{[({\bf k} + {\bf q}/2)^2+\Lambda^2][({\bf k} - {\bf q}/2)^2+\Lambda^2]}.
\label{eq:7}
\end{align}

Using $\epsilon_{1,2}(k)=v_xk_x\pm v_yk_y$, we can redefine ${\bf \tilde k}=(\tilde k_x,\tilde k_y)=(k_x+v_y/(2v_xq_y), k_y+v_x/(2v_yq_x))$, such that in Eq.\ (\ref{eq:6}) we have $\epsilon_1(k+q/2)=\epsilon_1(\tilde k)$ and $\epsilon_2(k-q/2)=\epsilon_2(\tilde k)$. This way all the $q$ dependences are absorbed into the Lorentzian multipliers:
\begin{align}
\Pi_{k_0}({\bf Q}+{\bf q})=&-\int\frac{d^2\tilde k}{4\pi^2}\frac{f(\epsilon_1(\tilde k))-f(\epsilon_2(\tilde k))}{\epsilon_1(\tilde k)-\epsilon_2(\tilde k)}\frac{\Lambda^{2}}{[{\tilde k}_x- v_yq_y/(2v_x) + {q_x}/2]^2+[{\tilde k}_y- v_xq_x/(2v_y) + {q_y}/2]^2+\Lambda^2}\nonumber\\
&\times\frac{\Lambda^2}{[{\tilde k}_x- v_yq_y/(2v_x) - {q_x}/2]^2+[{\tilde k}_y- v_xq_x/(2v_y) - {q_y}/2]^2+\Lambda^2},
\label{eq:8}
\end{align}
where, as before, we have summed over frequency. To obtain the quadratic coefficients in $q$'s, we can expand the Lorentian multipliers in $q_x$ and $q_y$'s. The result will then have the form $\Pi_{k_0}({\bf Q}+{\bf q})=\Pi_{k_0}({\bf Q},r)+D_x(r) q_x^2+D_y(r) q_y^2$, where we keep the $r=v_y/v_x$ dependence explicit, and 
\begin{align}
D_x\(\frac{v_y}{v_x}\)=&-\int\frac{d^2\tilde k}{8\pi^2}\frac{f(\epsilon_1(\tilde k))-f(\epsilon_2(\tilde k))}{\epsilon_1(\tilde k)-\epsilon_2(\tilde k)} \frac{d^2}{dq_x^2}\left\{\frac{\Lambda^{2}}{[{\tilde k}_x- v_yq_y/(2v_x) + {q_x}/2]^2+[{\tilde k}_y- v_xq_x/(2v_y) + {q_y}/2]^2+\Lambda^2}\right.\nonumber\\
&\times\left. \frac{\Lambda^2}{[{\tilde k}_x- v_yq_y/(2v_x) - {q_x}/2]^2+[{\tilde k}_y- v_xq_x/(2v_y) - {q_y}/2]^2+\Lambda^2}\right\}\bigg|_{q=0},\nonumber\\
D_y\(\frac{v_y}{v_x}\)=&-\int\frac{d^2\tilde k}{8\pi^2}\frac{f(\epsilon_1(\tilde k))-f(\epsilon_2(\tilde k))}{\epsilon_1(\tilde k)-\epsilon_2(\tilde k)} \frac{d^2}{dq_y^2}\left\{\frac{\Lambda^{2}}{[{\tilde k}_x- v_yq_y/(2v_x) + {q_x}/2]^2+[{\tilde k}_y- v_xq_x/(2v_y) + {q_y}/2]^2+\Lambda^2}\right.\nonumber\\
&\times\left. \frac{\Lambda^2}{[{\tilde k}_x- v_yq_y/(2v_x) - {q_x}/2]^2+[{\tilde k}_y- v_xq_x/(2v_y) - {q_y}/2]^2+\Lambda^2}\right\}\bigg|_{q=0}.
\end{align}
 Like we did before, $\Pi_{k_\pi}$ and $\Pi_{k_0}$ are related by $k_x\leftrightarrow k_y$ , hence $\Pi_{k_\pi}({\bf Q}+{\bf q})=\Pi_{k_0}({\bf Q},1/r)+D_x(1/r) q_x^2+D_y(1/r) q_y^2$. In the magnetic scenario, the spatial anisotropy of CDW correlation length $\xi_{\|}/\xi_{\perp}$ is given by the ratio of $q^2$ coefficients in $\Pi_{k_0}(q)\Pi_{k_\pi}(q)$, which means
\begin{align}
\frac{\xi_{\|}}{\xi_{\perp}}=\frac{\Pi_0({\bf Q},r)D_y(1/r)+\Pi_0({\bf Q},1/r)D_y(r)}{\Pi_0({\bf Q},r)D_x(1/r)+\Pi_0({\bf Q},1/r)D_x(r)}.
\end{align}
We used the experimental ratio $r=13.6$ and evaluated all the coefficients numerically, and we obtained $\xi_{\|}/\xi_{\perp}=27.3$. This ratio, however, is much larger than the experimental value $\sim 1.5$.

\section{Coupling between superconductivity and charge order}
We next compute the effect of feedback from the superconducting (SC) order parameter $\Delta_{sc}$ to the density and current component of the charge order (CO) parameter. We use two approaches. First, we compute the coupling between CO and SC perturbatively by evaluating the square diagrams (see Fig.2(c) in the main text). Second, we compute the particle-hole bubble in Nambu space, which naturally encapsulates the contribution from the SC order parameter $\Delta_{sc}$ into the normal and anomalous Green's functions, and at the end expand the result in terms of the SC order parameter. We show in Section \ref{Sec:can}, using both methods, which are equivalent by nature, that the leading contributions cancel for density component, while they add up for current component. This leads to a much weaker ``residual" feedback effect from the SC order parameter on the density component compared to the current component, which we compute in Section \ref{Sec:res}.

\subsection{Cancellation of the leading order coupling between density component of charge order and superconductivity}
\label{Sec:can}
We start by doing perturbation theory. At lowest order the coupling between $\Delta_{sc}$ and $\Delta$'s are given by square diagrams shown in Fig.\ 2(c) in the main text. Note that there are two types of such diagrams. In the first type, only the modulus of the CO parameter with a given center of mass momentum, $|\Delta_k^Q|^2$, is involved. In the second type, however, the CO parameters involved are of the form $\Delta_k^Q (\Delta_{-k}^Q)^*$. Therefore they correspond to the coefficients $\beta_{11}$ and $\beta_{12}$ in Eq.\ (3) of the main text respectively. 

Like the bubbles, the square diagrams should also be evaluated for $\Delta$'s with center of mass momenta at both $\pm k_0$ and $\pm k_\pi$. However, the ratio between $\Delta_{k_0}^Q$ and $\Delta_{k_\pi}^Q$ is fixed since they are coupled by the antiferromagnetic interaction~\cite{charge}. Moreover, by a simple transformation, one can show that the integrals for the square diagrams of the same type involving fermions near $k_0$ and $k_\pi$ are identical. Therefore, the contribution to the Free energy from these square diagrams involving $\Delta_{\pm k_\pi}$ just differ from those involving $\Delta_{\pm k_0}$ by a constant. For simplicity we will only consider the diagrams with $\Delta_{\pm k_0}$, and the extra constant can be easily absorbed into a rescaling of $\Delta_{sc}$.
Therefore, we have
\begin{align}
\beta_{11}&=-2T\sum_m\int \frac{d^2k}{4\pi^2} G_1(\omega_m,k)G_2(\omega_m,k)G_1(\omega_m,k)G_{-1}(-\omega_m,-k)\nonumber\\
\beta_{12}&=-T\sum_m \int \frac{d^2k}{4\pi^2}G_1(\omega_m,k)G_{-1}(-\omega_m,-k)G_2(\omega_m,k)G_{-2}(-\omega_m,-k),
\label{beta}
\end{align}
where $G_1(\omega_m,k)=1/[i\omega_m-\epsilon_1(k)]$.
Both integrals have been evaluated in Refs.\ \onlinecite{debanjan,coex} with linear dispersion $\epsilon_{1,2}$ and infinite limits on momentum integration. However, an overall sign of $\beta_{12}$ was overlooked. The revised results are of the form,
\begin{align}
\beta_{11}=&\frac{1}{32v_xv_y T}\[1+O\(\frac T\Lambda\)\],\nonumber\\
\beta_{12}=&-\frac{1}{32 v_x v_y T}\[1+O\(\frac T\Lambda\)\].
\label{eq2}
\end{align}
Plugging these into Eq.\ (4) of the main text, we immediately see that there is a cancellation of leading order ($\sim1/T$) contributions in the bi-quadratic coupling between the density component $\Delta_d$ and  $\Phi$, while the bi-quadratic coupling between the current component $\Delta_c$ remains $\sim 1/T$.

The same cancellation between leading order contributions can also be derived within the Nambu formalism in the SC state. For simplicity we focus on the region of hot spots 1,2,-1,-2 only. We introduce the Nambu spinors as, e.g.,
\begin{align}
\Psi_{1}^\dagger(k)=\(c_{1\uparrow}^\dagger(k)~~c_{-1\downarrow}(-k)\),~~\Psi_{2}^\dagger(k)=\(c_{2\uparrow}^\dagger(k)~~c_{-2\downarrow}(-k)\).
\end{align}
The SC gap $\Delta_{sc}$ can then be naturally included into the fermionic dispersion as
\begin{align}
{\cal{H}}_0+{\cal{H}}_{sc}= \Psi_{1}^\dagger(k) [\epsilon_1(k)\tau_z + \Delta_{sc}\tau_x]\Psi_{1}(k) + \Psi_{2}^\dagger (k) [\epsilon_2(k)\tau_z + \Delta_{sc}\tau_x]\Psi_{2}(k),
\end{align}
where $\tau_x$ and $\tau_z$ are Pauli matrices in Nambu space, and for simplicity we have taken $\Delta_{sc}$ to be real.
From the above, we can write down the Green's functions as,
\begin{align}
\mathcal{G}_{i}(\omega_m,{k})=&\left( \begin{array}{cc} 
G_{i}^0(\omega_m,{k}) & -F_{i}^0(\omega_m,{k})\\
-F_{i}^0(\omega_m,{k})  & -G_{i}^0(-\omega_m,{k})
\end{array}\right),\\
G_{i}^0(\omega_m,{k})=&-\frac{i\omega_m+\epsilon_{i}({k})}{\omega_m^2+E^2_{i}({k})},~F_{i}^0(\omega_m,k)=-\frac{\Delta_{sc}}{\omega_m^2+E^2_{i}({k})},~\text{where}\\
E_{i}^2({k})=&\epsilon_{i}^2({k})+|\Delta_{sc}|^2.
\end{align}

 We now couple the fermions $\Psi$'s to charge order parameters $\Delta_d$ and $\Delta_c$. Using the fact that $\Delta_{d(c)}$ is the even (odd) combination of the CO parameters $\Delta_{\pm k_0}^Q$, we have
\begin{align}
\mathcal{H}_{\Delta}= \Delta_d(\Psi_{1}^\dagger(k)\tau_z\Psi_{2}(k))+\Delta_c(\Psi_{1}^\dagger(k)\tau_0\Psi_{2}(k)),
\end{align}
where $\tau_0$ is an identity matrix.
We evaluate the bubbles for $\Delta_{d,c}$, and the coupling between $\Delta_{d,c}$ and $\Delta_{sc}$ can be obtained by expanding these bubbles in powers of $\Delta_{sc}$. 

For $\Delta_d$, we find
\begin{align}
\bar \Pi_{k_0}^{d}({\bf Q}_y)=&-T\sum_m\int\frac{d^2k}{(2\pi)^2} \Tr[\mathcal{G}_{1}(\omega_m,k)\cdot\tau_z\cdot\mathcal{G}_{2}(\omega_m,k)\cdot\tau_z]  \nonumber\\
=&2T\sum_m\int\frac{d^2k}{(2\pi)^2}  \frac{\omega_m^2-\epsilon_1(k)\epsilon_2(k)+|\Delta_{sc}|^2}{[\omega_m^2+E_{1}^2(k)][\omega_m^2+E_{2}^2(k)]}.
\label{pd}
\end{align}
At this step if we expand in the integrand with respect to $\Delta_{sc}$ to quadratic order, the coefficients we get are precisely $\beta_{11}+\beta_{12}$. As we have seen, both $\beta$'s scale as $\sim 1/T$ and cancel out. Here we do something slightly different to show the same cancellation. Namely, we take the $T\to 0$ limit and replace the Matsubara sum with an integral, then
\begin{align}
\bar \Pi_{k_0}^{d}({\bf Q}_y)=&\int\frac{d\omega_m}{\pi}\int\frac{d^2k}{(2\pi)^2}  \frac{\omega_m^2-\epsilon_1(k)\epsilon_2(k)+|\Delta_{sc}|^2}{[\omega_m^2+E^2_{1}(k)][\omega_m^2+E^2_{2}(k)]}\nonumber\\
=&\int\frac{d^2k}{(2\pi)^2} \frac{E_{1}(k)E_{2}(k)-\epsilon_1(k)\epsilon_2(k)+|\Delta_{sc}|^2}{E_{1}(k)E_{2}(k)[E_{1}(k)+E_{2}(k)]}.
\end{align}
To compute the effect of $\Delta_{sc}$, we subtract from the above integral its $\Delta_{sc}$-independent part, namely 2$\Pi_{k_0}({\bf Q})$.
\begin{align}
\delta\bar \Pi_{k_0}^{d}({\bf Q}_y)=\int\frac{d^2k}{(2\pi)^2} \[\frac{E_{1}(k)E_{2}(k)-\epsilon_1(k)\epsilon_2(k)+|\Delta_{sc}|^2}{E_{1}(k)E_{2}(k)[E_{1}(k)+E_{2}(k)]}-\frac{|\epsilon_{1}(k)\epsilon_{2}(k)|-\epsilon_1(k)\epsilon_2(k)}{|\epsilon_{1}(k)\epsilon_{2}(k)|(|\epsilon_{1}(k)|+|\epsilon_{2}(k)|)}\].
\label{dP}
\end{align}
To obtain leading order contribution, which we know comes from the infrared, we can evaluate Eq.\ (\ref{dP}) with linear dispersion $\epsilon_1,\epsilon_2$ and infinite integration limits. A little trick here is to transform the integration variables from $k=(k_x, k_y)$ to $(\epsilon_1,\epsilon_2)$. The Jacobian of this transformation is $1/(2v_xv_y)$. It is then clear that the $\epsilon_1\epsilon_2$ terms in both numerators are odd and hence can be eliminated. We regroup $\delta\bar \Pi_{k_0}^d$ into two terms,
\begin{align}
\delta \bar\Pi_{k_0}^d({\bf Q}_y)=&\int\frac{d\epsilon_1 d\epsilon_2}{8\pi^2 v_x v_y} \[\frac{|\Delta_{sc}|^2}{E_{1}E_{2}(E_{1}+E_{2})} + \(\frac{1}{E_1+E_2}-\frac{1}{|\epsilon_1|+|\epsilon_2|}\) \] \nonumber\\
=&\frac{|\Delta_{sc}|}{8v_xv_y}\[\int_{-\infty}^{\infty}\frac{d\varepsilon_1 d\varepsilon_2}{\pi^2}  \frac{1}{\sqrt{1+\varepsilon^2_{1}}\sqrt{1+\varepsilon^2_{2}}(\sqrt{1+\varepsilon_1^{2}}+\sqrt{1+\varepsilon^2_{2}})}\right.\nonumber\\ 
&~~~~~~~+\left.\int_{-\infty}^{\infty}\frac{d\varepsilon_1 d\varepsilon_2}{\pi^2} \(\frac{1}{\sqrt{1+\varepsilon_1^2}+\sqrt{1+\varepsilon_2^2}}-\frac{1}{|\varepsilon_1|+|\varepsilon_2|}\) \]
\end{align}
where we have defined $\varepsilon_{1,2}=\epsilon_{1,2}/|\Delta_{sc}|$.
Naively, this means the leading order renormalization to the bubble for charge order is of order $\sim|\Delta_{sc}|$. This $|\Delta_{sc}|$ scaling is nothing but the reminiscence of the $|\Delta_{sc}|^2/T$ at zero temperature. However, a direct evaluation of the integrals yields,
\begin{align}
\int_{-\infty}^{\infty}\frac{d\varepsilon_1 d\varepsilon_2}{\pi^2}  \frac{1}{\sqrt{1+\varepsilon^2_{1}}\sqrt{1+\varepsilon^2_{2}}(\sqrt{1+\varepsilon_1^{2}}+\sqrt{1+\varepsilon^2_{2}})}=1, \nonumber\\
\int_{-\infty}^{\infty}\frac{d\varepsilon_1 d\varepsilon_2}{\pi^2} \(\frac{1}{\sqrt{1+\varepsilon_1^2}+\sqrt{1+\varepsilon_2^2}}-\frac{1}{|\varepsilon_1|+|\varepsilon_2|}\)=-1.
\label{eq:33}
\end{align}
 Therefore, at leading order, $\delta\bar\Pi_{k_0}^d({\bf Q}_y)=0$. This reproduces our previous result on the cancellation of leading order couplings between $\Delta_d$ and $\Delta_{sc}$.

Now we repeat the above procedure for $\Delta_c$, the odd component of $\Delta_{k_0}^Q$. The bubble is given by
\begin{align}
\bar \Pi_{k_0}^{c}({\bf Q}_y)=&-T\sum_m\int\frac{d^2k}{(2\pi)^2} \Tr[\mathcal{G}_{1}(\omega_m,k)\cdot\tau_0\cdot\mathcal{G}_{2}(\omega_m,k)\cdot\tau_0]  \nonumber\\
=&2T\sum_m\int\frac{d^2k}{(2\pi)^2}  \frac{\omega_m^2-\epsilon_1(k)\epsilon_2(k)-|\Delta_{sc}|^2}{[\omega_m^2+E^2_{1}(k)][\omega_m^2+E^2_{2}(k)]}.
\label{pc}
\end{align}
The only difference between Eq.\ (\ref{pc}) and (\ref{pd}) is the sign in front of $|\Delta_{sc}|^2$. Subtracting the $\Delta_{sc}$-independent part and rewriting, we obtain
\begin{align}
\delta \bar\Pi_{k_0}^c({\bf Q})=&\int\frac{d\epsilon_1 d\epsilon_2}{8\pi^2 v_x v_y} \[-\frac{|\Delta_{sc}|^2}{E_{1}E_{2}(E_{1}+E_{2})} + \(\frac{1}{E_1+E_2}-\frac{1}{|\epsilon_1|+|\epsilon_2|}\) \] \nonumber\\
=&\frac{|\Delta_{sc}|}{8v_xv_y}\[-\int_{-\infty}^{\infty}\frac{d\varepsilon_1 d\varepsilon_2}{\pi^2}  \frac{1}{\sqrt{1+\varepsilon^2_{1}}\sqrt{1+\varepsilon^2_{2}}(\sqrt{1+\varepsilon_1^{2}}+\sqrt{1+\varepsilon^2_{2}})}\right.\nonumber\\ 
&~~~~~~~+\left.\int_{-\infty}^{\infty}\frac{d\varepsilon_1 d\varepsilon_2}{\pi^2} \(\frac{1}{\sqrt{1+\varepsilon_1^2}+\sqrt{1+\varepsilon_2^2}}-\frac{1}{|\varepsilon_1|+|\varepsilon_2|}\) \]\nonumber\\
=&-\frac{|\Delta_{sc}|}{4v_xv_y}.
\end{align}
where in the last line we have inserted the value of the two integrals in Eqs.\ (\ref{eq:33}) in the bracket.

\subsection{The residual coupling between density component of charge order and superconductivity}
\label{Sec:res}
Since the leading order coupling between $\Delta_d$ and $\Delta_{sc}$ vanishes, we need to go beyond leading order to compute their ``residual" coupling, and we need to verify that this coupling is indeed repulsive. This can be done if one explicitly keeps the upper momentum cutoff $\Lambda$ in evaluating $\beta_{11}$ and $\beta_{12}$. However the analytical computation for the $\beta$'s with a circular hard or soft momentum cutoff is more difficult than the computation of the bubbles in the normal state. Here we consider a case when a cutoff $\bar\Lambda_1,~\bar\Lambda_2$ is imposed on $\epsilon_1$ and $\epsilon_2$ respectively. 
We have evaluated the $\beta$'s with more realistic cutoff's numerically and have found that the result is qualitatively the same.

We start from Eq.\ (\ref{beta}). After converting the integration variable from $(k_x,k_y)$ to $(\epsilon_1,\epsilon_2)$, and imposing a integration cutoff as described above, we have,
\begin{align}
\beta_{11}=&2T\sum_m\int_{-\bar\Lambda_1}^{\bar \Lambda_1}\int_{-\bar\Lambda_2}^{\bar \Lambda_2}\frac{d\epsilon_1 d\epsilon_2}{8\pi^2v_xv_y}\frac{\omega_m^2}{(\omega_m^2+\epsilon_1^2)^2}\frac{1}{\omega_m^2+\epsilon_2^2},\nonumber\\
\beta_{12}=&-T\sum_m\int_{-\bar\Lambda_1}^{\bar \Lambda_1}\int_{-\bar\Lambda_2}^{\bar \Lambda_2}\frac{d\epsilon_1 d\epsilon_2}{8\pi^2v_xv_y}\frac{1}{\omega_m^2+\epsilon_1^2}\frac{1}{\omega_m^2+\epsilon_2^2}.
\end{align}
We integrate over $\epsilon_{1,2}$ first,

\begin{align}
\beta_{11}=&\frac{T}{2\pi^2 v_x v_y}\sum_m\[\frac{1}{|\omega_m|}\tan^{-1}\(\frac{\bar\Lambda_1}{|\omega_m|}\)+\frac{\bar\Lambda_1}{\bar\Lambda_1^2+\omega_m^2}\]\[\frac{1}{|\omega_m|}\tan^{-1}\(\frac{\bar\Lambda_2}{|\omega_m|}\)\],\nonumber\\
\beta_{12}=&\frac{-T}{2\pi^2 v_x v_y}\sum_m\[\frac{1}{|\omega_m|}\tan^{-1}\(\frac{\bar\Lambda_1}{|\omega_m|}\)\]\[\frac{1}{|\omega_m|}\tan^{-1}\(\frac{\bar\Lambda_2}{|\omega_m|}\)\],\end{align}
The major contributions to both Matsubara summations come from the infrared, therefore we can expand the summand as Laurent series of $|\omega_m|/\bar\Lambda_{1,2}$. Up to next-leading order, we have
\begin{align}
 \beta_{11}\approx&\frac{T}{2\pi v_xv_y}\sum_m\[\(\frac{\pi}{2}\)^2\frac{1}{|\omega_m|^2}-\frac{\pi}{2\bar\Lambda_2|\omega_m|}+\cdots\]\nonumber\\
  \beta_{12}\approx&\frac{-T}{2\pi v_xv_y}\sum_m\[\(\frac{\pi}{2}\)^2\frac{1}{|\omega_m|^2}-\frac{\pi}{2|\omega_m|}\(\frac{1}{\bar\Lambda_1}+\frac{1}{\bar\Lambda_2}\)+\cdots\].
 \end{align}
where $\cdots$ stands for terms that are not divergent in $1/|\omega_m|$. 

The first terms in both summands give the same results as shown in Eq.\ (\ref{eq2}), which cancel out in $(\beta_{11}+\beta_{12})$. Therefore, the bi-quadratic coupling between $\Delta_d$ and $\Delta_{sc}$ is
\begin{align}
\beta_{11}+\beta_{12}=\frac{1}{4\pi \bar\Lambda_1v_xv_y}\log\(\frac{\bar\Lambda}{T} \),
\label{leftover}
 \end{align}
 while the bi-quadratic coupling between $\Delta_c$ and $\Delta_{sc}$, at leading order, is
\begin{align}
\beta_{11}-\beta_{12}=\frac{1}{16v_xv_yT}.
\end{align}
Eq.\ (\ref{leftover}) is consistent with the experiment~\cite{damascelli_new}, since it gives rise to a weak but finite suppression of the CDW correlation length upon entering the superconducting phase.

\end{widetext}
\end{document}